\documentclass{article}

\usepackage[preprint]{neurips_2025}

\usepackage[utf8]{inputenc}
\usepackage[T1]{fontenc}
\usepackage{hyperref}
\usepackage{url}
\usepackage{booktabs}
\usepackage{amsfonts}
\usepackage{amsmath}
\usepackage{amssymb}
\usepackage{graphicx}
\usepackage{xcolor}
\usepackage{subcaption}
\usepackage{multirow}
\usepackage{makecell}
\usepackage{enumitem}

\newcommand{\tocs}{\textsc{ToCS}}
\newcommand{\tos}{\textsc{ToS}}
\newcommand{\pms}[1]{\,{\scriptsize$\pm$\,#1}}

\title{Theory of Code Space: Do Code Agents Understand Software Architecture?}

\author{
  Grigory Sapunov\thanks{Work in progress. All LLM results are preliminary (single run, single prompt, one pattern); model comparison at scale is future work. Collaborations welcome.} \\
  Intento \\
  \texttt{gs@inten.to}
}

\begin{document}

\maketitle

% ============================================================================
\begin{abstract}
AI code agents excel at isolated tasks yet struggle with multi-file software engineering requiring architectural understanding. We introduce \textbf{Theory of Code Space} (\tocs{}), a benchmark that evaluates whether agents can construct, maintain, and update coherent \emph{architectural beliefs} during codebase exploration. Agents explore procedurally generated codebases under partial observability---opening files under a budget---and periodically externalize their belief state as structured JSON, producing a time-series of architectural understanding. Three findings emerge from experiments with four baselines and six frontier LLMs. First, the Active-Passive Gap is \emph{model-dependent}: one model builds better maps through active exploration than from seeing all files at once, while another shows the opposite---revealing that active exploration is itself a non-trivial capability absent from some models. Second, retaining structured belief maps in context acts as self-scaffolding for some models but not others, showing that the mechanism is model-dependent. Third, \emph{belief state maintenance} varies dramatically: a smaller model maintains perfectly stable beliefs across probes while its larger sibling suffers catastrophic belief collapse---forgetting previously-discovered components between probes. We release \tocs{} as open-source software.\footnote{Code: \url{https://github.com/che-shr-cat/tocs}}
\end{abstract}

% ============================================================================
\section{Introduction}
\label{sec:intro}

Large language models achieve remarkable scores on code generation benchmarks~\citep{chen2021codex, austin2021program}, creating an expectation that they possess deep understanding of software architecture. Yet practitioners report a persistent gap: models that solve HumanEval problems with ease produce incoherent results when modifying real codebases with dozens of interdependent modules~\citep{jimenez2024swebench}. This gap has resisted satisfying explanation.

Recent work on spatial reasoning offers a compelling diagnostic framework. \citet{zhang2026theoryofspace} introduced the \emph{Theory of Space} (\tos{}) benchmark, demonstrating that multimodal models fail to maintain coherent internal ``cognitive maps'' when actively exploring partially observable environments. They identified two core phenomena: an \textbf{Active-Passive Gap} (models degrade when they must gather information themselves rather than receiving it upfront) and \textbf{Belief Inertia} (models cannot update their spatial beliefs when the environment changes). We hypothesize that the same latent state maintenance failures explain the struggles of code agents. The term ``cognitive map'' traces to \citet{tolman1948cognitive}; in program comprehension, \citet{pennington1987stimulus} and \citet{vonmayrhauser1995program} showed that developers build layered mental models of code---our benchmark operationalizes and measures this process for AI agents.

We propose \textbf{Theory of Code Space} (\tocs{}), a benchmark that transplants this diagnostic framework to software engineering. Rather than grid worlds, our ``environment'' consists of procedurally generated codebases with controlled architectural structure. Rather than allocentric spatial maps, our ``cognitive maps'' are structured representations of module dependencies, typed edges, cross-cutting invariants, and design intent. The agent explores under partial observability---opening files one at a time under a budget---and must externalize its architectural belief as structured JSON at regular intervals.

Beyond the spatial analogy, we introduce \textbf{Architectural Constraint Discovery}---a dimension absent from spatial benchmarks. In code, architectural decisions encode \emph{checkable constraints}: a forbidden dependency enforces a service boundary; a validation chain ensures data integrity. We plant explicit, verifiable constraints in our generated codebases and measure whether agents can discover them.

Our contributions: (1)~the \tocs{} benchmark framework for active architectural belief construction in code (\S\ref{sec:benchmark}); (2)~a procedural codebase generator with four typed edge categories and planted invariants (\S\ref{sec:generation}); (3)~pilot experiments with four baselines and six LLMs showing a wide capability spread and three surprising findings---a model-dependent Active-Passive Gap, model-dependent self-scaffolding effects, and dramatic belief state instability differences across models (\S\ref{sec:results}--\ref{sec:probe_ablation}); (4)~an open benchmark released for community evaluation (\S\ref{sec:discussion}).

% ============================================================================
\section{Related Work}
\label{sec:related}

\paragraph{Code Generation Benchmarks.}
SWE-bench~\citep{jimenez2024swebench} evaluates bug-fixing in real GitHub repositories but measures patch correctness, not evolving architectural understanding. ContextBench~\citep{li2026contextbench} advances this by logging agent trajectories and scoring context retrieval precision/recall against human-verified gold contexts---measuring \emph{what the agent looked at} but not \emph{what the agent believes about architecture}. SWE-ContextBench~\citep{zhu2026swecontextbench} tests cross-task experience reuse. RepoBench~\citep{liu2024repobench} and LoCoBench-Agent~\citep{qiu2025locobench} evaluate repo-level completion and long-context interaction respectively, with LoCoBench observing a ``comprehension-efficiency trade-off'' that we formalize as the Active-Passive Gap. RefactorBench~\citep{masai2025refactorbench} targets multi-file refactoring but evaluates output correctness without probing belief state. None of these benchmarks require agents to \emph{externalize} a revisable architectural belief state, impose exploration budgets, or score against typed dependency graphs with planted constraints. We note that our own belief externalization finding (\S\ref{sec:results}) reveals a complementary limitation: even when agents \emph{do} externalize beliefs, what they report may not reflect what they internally represent.

\paragraph{Code Understanding Tools.}
CodePlan~\citep{bairi2024codeplan} augments LLMs with static-analysis dependency graphs for change propagation---an engineering solution that validates our premise: agents need architectural maps but currently cannot build them alone. Aider's RepoMap~\citep{gauthier2024aider} uses tree-sitter parsing and PageRank to construct repository context. Code World Models~\citep{fair2025cwm} train LLMs on execution traces to improve code reasoning. \tocs{} provides the diagnostic benchmark to test whether such approaches actually improve architectural belief quality.

\paragraph{Spatial Reasoning.}
Theory of Space~\citep{zhang2026theoryofspace} demonstrated Active-Passive Gap and Belief Inertia in grid-world environments with multimodal models. Related spatial benchmarks include SpatialVLM~\citep{spatialvlm2024}, Habitat~\citep{habitat2019}, and OpenEQA~\citep{openeqa2024}, which evaluate spatial reasoning and embodied question answering but do not require constructing revisable belief states. We transplant this framework to code, adding architectural constraint discovery (absent from spatial domains where object placement ``why'' is rarely meaningful). TOM-SWE~\citep{zhou2025tomswe} models the \emph{user's} mental state in SWE agents---complementary to our focus on the agent's belief about the \emph{codebase}.

% ============================================================================
\section{Theory of Code Space}
\label{sec:benchmark}

\subsection{Problem Formulation}

We formalize codebase understanding as \emph{belief construction} over a latent architectural state. Let $S$ denote the ground-truth architecture, comprising a typed dependency graph $G = (V, E)$ with edge types $\tau \in \{\textsc{imports}, \textsc{calls\_api}, \textsc{data\_flows\_to}, \textsc{registry\_wires}\}$, a set of cross-module invariants $\mathcal{I}$, and exported contracts $\mathcal{C}$. At time $t$, the agent has history $h_t = (o_{0:t}, a_{0:t})$ and we evaluate three operations on its belief $\mathcal{M}_t(S)$.  Since $\mathcal{M}_t$ is a latent internal state, we measure it through \emph{probing}: periodically asking the agent to externalize its belief as structured JSON (\S\ref{sec:probing}). We evaluate:

\begin{description}[nosep]
    \item[Construct] Build $\mathcal{M}_t$ from partial observations gathered through active exploration.
    \item[Revise] Update $\mathcal{M}_t \to \mathcal{M}_{t+\Delta t}$ when the environment mutates ($S \to S'$) and the agent encounters evidence of the change.
    \item[Exploit] Use $\mathcal{M}_t$ to perform a downstream engineering task correctly (v1.0; v0.1 uses counterfactual probes as a proxy).
\end{description}
\noindent This paper evaluates \textbf{Construct}; Revise and Exploit are implemented in the framework but reserved for future evaluation (\S\ref{sec:roadmap}).

\subsection{Environment and Action Space}

The agent interacts with the codebase through five actions with \textbf{fixed tool semantics}:
\begin{itemize}[nosep]
    \item \texttt{LIST}($d$): Filenames in directory $d$ (no contents, no recursion).
    \item \texttt{OPEN}($f$): Full contents of file $f$. Costs 1 action.
    \item \texttt{SEARCH}($q$): Matching filepaths + line numbers. \textbf{No content snippets.} Costs 1 action.
    \item \texttt{INSPECT}($f$, $s$): Type signature + docstring of symbol $s$. No body. Costs 1 action.
    \item \texttt{DONE}(): Terminate. Costs 0 actions.
\end{itemize}
The agent operates under a budget $B$ (default $B{=}20$). Critically, \texttt{SEARCH} returns only locations, never content---ensuring that architectural understanding requires deliberate \texttt{OPEN} decisions. \texttt{INSPECT} provides a lightweight discovery channel: docstrings often describe module purpose and relationships (e.g., ``wraps stage X''), allowing agents to gather architectural hints at the cost of one action without reading full file contents.

\subsection{Cognitive Map Probing}
\label{sec:probing}

Every $K{=}3$ actions, the harness interrupts the agent with a structured prompt requesting it to externalize its current architectural belief as JSON. Critically, \textbf{probing is free}: it does not consume an action from the budget $B$, and the agent retains the same remaining budget before and after a probe. This ensures that measurement does not interfere with exploration strategy---agents are evaluated on their understanding, not on a trade-off between exploring and reporting.

The externalized cognitive map $\hat{\mathcal{M}}_t$ contains: (1) \textbf{component beliefs} with status (observed/inferred/unknown), purpose, exported symbols with typed signatures, and typed dependency edges with per-element confidence; (2) \textbf{invariant beliefs} with structured canonical form (\texttt{type, src, dst, via, pattern}) and evidence pointers; (3) \textbf{uncertainty tracking}---explicit list of unexplored regions. The resulting time-series of maps (one every $K$ steps) captures \emph{how} understanding develops, not just the final state.

\subsection{Evaluation Modes}

\tocs{} decomposes the Active-Passive Gap through four conditions:
\begin{itemize}[nosep]
    \item \textbf{Active}: Agent chooses actions under budget $B$.
    \item \textbf{Passive-Full}: Agent receives the entire codebase; probed once.
    \item \textbf{Passive-Oracle}: Agent receives $B$ files selected by oracle (maximum ground-truth connectivity); one file per step, probed every $K$.
    \item \textbf{Passive-Replay}: Agent receives the \emph{exact observation trace} from a prior active run, without making decisions. Each observation counts as one step.
\end{itemize}

This decomposes: $\text{APG}_\text{total}$ (passive-full $-$ active) into $\text{APG}_\text{selection}$ (passive-oracle $-$ active, the cost of choosing \emph{which} files) and $\text{APG}_\text{decision}$ (passive-replay $-$ active, the cost of deciding \emph{what to do} with observations).

\noindent We report APG decomposition results for GPT-5.3-Codex and Gemini~2.5~Flash in \S\ref{sec:apg}.

\subsection{Architectural Constraint Discovery}

Beyond structural mapping, we probe whether agents discover planted architectural constraints:
\begin{itemize}[nosep]
    \item \textbf{Forbidden dependency}: ``Module $A$ must not import module $C$ directly.''
    \item \textbf{Interface-only access}: ``Module $X$ must access $Y$ only through interface $Z$.''
    \item \textbf{Validation chain}: ``Data must pass through validation before reaching module $W$.''
\end{itemize}
Each constraint has a \textbf{discoverability requirement}: test evidence, structural patterns, or documentation in the codebase. Scored via counterfactual multiple-choice probes (``Which change would violate an architectural constraint?'').

% ============================================================================
\section{Procedural Codebase Generation}
\label{sec:generation}

\subsection{Architecture Grammar}

Our v0.1 generator produces \textbf{Pipeline} architecture codebases with controlled anti-triviality measures. A \texttt{PipelineTemplate} grammar defines:

\begin{itemize}[nosep]
    \item \textbf{Domain pools}: Three domains (data ETL, log processing, text processing), each with 8 semantically coherent processing stages. Medium complexity selects 6--8 stages.
    \item \textbf{Module roles}: Infrastructure (models, base class, config, exceptions, registry), stages (processing steps implementing a common ABC), adapters (wrapping stages with additional behavior), middleware (cross-cutting decorators), utilities, and legacy/distractor modules.
    \item \textbf{Anti-triviality}: (1) \emph{Registry wiring}---stages connected via config, not direct imports; (2) \emph{Adapter indirection}---ABC interface layer; (3) \emph{Distractor modules}---files not in the main pipeline; (4) \emph{Neutral naming}---\texttt{mod\_a.py}, not \texttt{extract.py}; (5) \emph{Hidden invariants}---constraints discoverable only by reading function bodies or tests.
\end{itemize}

\subsection{Four Edge Types}

Each generated codebase contains edges of four types, reflecting different \emph{methods of discovery}:
\begin{itemize}[nosep]
    \item \textsc{imports} ($\sim$67\%): Python \texttt{import} statements. Discoverable by AST parsing.
    \item \textsc{calls\_api} ($\sim$17\%): Runtime function calls between modules. Requires reading function bodies; docstrings may provide hints (e.g., ``delegates to module X'').
    \item \textsc{registry\_wires} ($\sim$9\%): Config-driven connections where a registry module dynamically loads stages via \texttt{importlib} based on names listed in a JSON config file. No static \texttt{import} statement exists. Requires reading both the config and the registry's loading logic.
    \item \textsc{data\_flows\_to} ($\sim$7\%): Data dependency where one module's output feeds another's input. Requires understanding orchestration logic; docstrings on the orchestrator may describe the data flow.
\end{itemize}

These proportions emerge from the Pipeline template: static imports dominate, while semantic edges arise only at architectural boundaries. Crucially, roughly one-third of edges are \emph{invisible to import-following}, creating a meaningful gap between syntactic analysis and semantic understanding.

\subsection{Invariant Planting}

Each codebase contains 15--16 planted constraints across five types: \textsc{boundary} (forbidden dependencies), \textsc{dataflow} (required processing chains), \textsc{interface} (access-only-through-ABC), \textsc{invariant} (naming/structural conventions), and \textsc{purpose} (design rationales). Every constraint has a \emph{structured canonical form} with five fields for machine-comparable scoring: \texttt{type} (constraint category), \texttt{src} and \texttt{dst} (the modules involved), \texttt{via} (the intermediary or mechanism, e.g., an ABC interface), and \texttt{pattern} (the structural rule, e.g., ``no direct import''). For example, a boundary constraint might be \texttt{(boundary, mod\_a, mod\_c, base.StageBase, ``must access only through ABC'')}. Each constraint has at least one evidence source (test file, structural pattern, or documentation).

\subsection{Generation Statistics}

\begin{table}[!ht]
\centering
\caption{Generated codebase characteristics (medium complexity, 3 seeds).}
\label{tab:codebases}
\begin{tabular}{@{}lccc@{}}
\toprule
\textbf{Metric} & \textbf{Seed 42} & \textbf{Seed 123} & \textbf{Seed 999} \\
\midrule
Modules & 27 & 30 & 27 \\
Total edges & 70 & 84 & 70 \\
\quad \textsc{imports} & 47 (67\%) & 56 (67\%) & 47 (67\%) \\
\quad \textsc{calls\_api} & 12 (17\%) & 15 (18\%) & 12 (17\%) \\
\quad \textsc{data\_flows\_to} & 5 (7\%) & 6 (7\%) & 5 (7\%) \\
\quad \textsc{registry\_wires} & 6 (9\%) & 7 (8\%) & 6 (9\%) \\
Invariants & 15 & 16 & 15 \\
Sub-packages & 5 & 5 & 5 \\
\bottomrule
\end{tabular}
\end{table}

% ============================================================================
\section{Metrics}
\label{sec:metrics}

\paragraph{Dependency F1.}
Compare predicted edges from the cognitive map against ground truth. An edge matches if (source, target, type) all agree via exact string matching. Targets must be specific file paths: a directory-level edge (e.g., \texttt{registry.py $\to$ stages/}) does not match file-level ground truth (e.g., \texttt{registry.py $\to$ stages/mod\_a.py}). Similarly, symbol-qualified targets (e.g., \texttt{base.py::StageBase} instead of \texttt{base.py}) do not match even when the file path portion is correct. This strict matching is deliberate---our ground truth is component-level (one file = one node)---but means that models reporting correct relationships at the wrong granularity or format receive no credit. We report precision, recall, and F1 separately.

\paragraph{Invariant F1.}
Match agent-discovered constraints against planted constraints via \emph{structured form} comparison on \texttt{(type, src, dst, via)}, not text similarity. \textbf{Strict} matching requires exact agreement on all four fields. \textbf{Relaxed} matching normalizes paths (directory prefix stripped), treats empty ground-truth fields as wildcards, and uses greedy 1-to-1 assignment---accommodating models that identify the correct constraint type and endpoints but use slightly different path formats. We report relaxed F1 in the main tables; strict scores are near-zero for all models due to compounding format mismatches.

\paragraph{Confidence Calibration.}
Expected Calibration Error (ECE) between agent-stated per-edge confidence and actual correctness, binned into 5 confidence intervals.

\paragraph{Two Efficiency Curves.}
\emph{Action-efficiency}: AUC(F1 vs.\ total actions)---the headline metric. \emph{Observation-efficiency}: AUC(F1 vs.\ \texttt{OPEN} count)---diagnostic, isolating information gain per file opened. Belief is piecewise-constant between probes; integration is trapezoidal.

\paragraph{Active-Passive Gap.}
For each metric $m$: $\text{APG}_m = m_\text{passive} - m_\text{active}$, decomposed into selection and decision components as described in \S\ref{sec:benchmark}.

\paragraph{Belief Revision Score (BRS).}
After a mutation at time $t_m$: $\text{BRS} = |\text{correctly updated}| / |\text{affected elements}|$. Decomposed into \emph{Inertia-proper} (correctly-believed elements not updated after evidence) and \emph{Impact-discovery} (missing elements newly found). A \emph{sham condition} (evidence without actual change) measures Gullibility. (BRS evaluation is implemented but reserved for future work.)

% ============================================================================
\section{Experiments}
\label{sec:results}

We evaluate four rule-based exploration strategies and six frontier LLM agents from three providers on three generated codebases (seeds 42, 123, 999) under identical conditions: budget $B{=}20$, probe interval $K{=}3$.

\subsection{Methods}

\paragraph{Rule-based baselines.}
\begin{itemize}[nosep]
    \item \textbf{Oracle}: Outputs the ground-truth graph directly. Upper bound (F1 = 1.0).
    \item \textbf{Config-Aware}: Lists all directories, opens config/registry files first, parses module references, then follows imports BFS.
    \item \textbf{Random}: Lists root, then opens files uniformly at random until budget exhausted. Builds map from observed imports.
    \item \textbf{BFS-Import}: Opens files breadth-first following import chains.
\end{itemize}
All baselines build cognitive maps from AST-parsed imports and config references.

\paragraph{LLM agents.}
\begin{itemize}[nosep]
    \item \textbf{GPT-5.3-Codex}: OpenAI's code-specialized reasoning model (\texttt{gpt-5.3-codex}). A reasoning model (temperature fixed at 1 per API constraint).
    \item \textbf{Claude Sonnet 4.6}: Anthropic's frontier coding model (\texttt{claude-sonnet-4-6}).
    \item \textbf{Gemini 2.5 Flash}: Google's reasoning-optimized flash model (\texttt{gemini-2.5-flash}). A ``thinking'' model with extended internal reasoning.
    \item \textbf{Gemini 2.5 Pro}: Google's reasoning-optimized pro model (\texttt{gemini-2.5-pro}). Larger sibling of 2.5~Flash.
    \item \textbf{Gemini 3 Flash}: Google's frontier flash model (\texttt{gemini-3-flash-preview}).
    \item \textbf{Gemini 3.1 Pro}: Google's most capable model (\texttt{gemini-3.1-pro-preview}). The flagship of the Gemini~3 family.
\end{itemize}
All models receive identical prompts: a system prompt describing the partial observability environment and available actions, per-turn action prompts showing remaining budget and opened files, and structured JSON probe prompts with schema, per-type field semantics, and worked examples for each invariant type (Appendix~\ref{app:prompts}). Relative to an earlier prompt version, the current probe includes explicit edge type decision rules and pairwise invariant decomposition instructions. Temperature is set to 0 except for GPT-5.3-Codex (reasoning model constraint). All probe responses were parsed successfully via the deterministic repair parser (zero format failures across all runs).

\subsection{Results}

\begin{table}[!ht]
\centering
\caption{Performance over 3 codebases (mean\pms{half-range}). Baselines use AST parsing; LLM agents use prompted JSON externalization with improved probe prompts (per-type field semantics, pairwise invariant decomposition). Best non-Oracle in \textbf{bold}. Inv~F1 uses relaxed matching (\S\ref{sec:probing}).}
\label{tab:results}
\begin{tabular}{@{}lccccccc@{}}
\toprule
\textbf{Method} & \textbf{Type} & \textbf{Dep F1} & \textbf{Precision} & \textbf{Recall} & \textbf{AUC} & \textbf{Inv F1} & \textbf{Files} \\
\midrule
Oracle          & baseline & 1.000 & 1.000 & 1.000 & ---   & --- & --- \\
Config-Aware    & baseline & 0.577\pms{.014} & 0.736 & 0.475 & 0.212\pms{.007} & 0.000 & 13 \\
Random          & baseline & 0.538\pms{.028} & 1.000 & 0.368 & 0.142\pms{.008} & 0.000 & 13 \\
BFS-Import      & baseline & 0.293\pms{.060} & 1.000 & 0.173 & 0.079\pms{.003} & 0.000 & 13 \\
\midrule
GPT-5.3-Codex   & LLM     & \textbf{0.676}\pms{.059} & 0.782 & \textbf{0.597} & 0.306\pms{.018} & 0.739\pms{.067} & 15 \\
Claude Sonnet 4.6 & LLM   & 0.664\pms{.026} & \textbf{0.983} & 0.502 & \textbf{0.350}\pms{.005} & \textbf{0.778}\pms{.025} & 13 \\
Gemini 2.5 Flash & LLM    & 0.470\pms{.100} & 0.642 & 0.373 & 0.259\pms{.037} & 0.517\pms{.154} & 14 \\
Gemini 3.1 Pro  & LLM     & 0.321\pms{.128} & 0.764 & 0.212 & 0.140\pms{.049} & 0.423\pms{.209} & 11 \\
Gemini 2.5 Pro  & LLM     & 0.242\pms{.034} & \textbf{1.000} & 0.138 & 0.205\pms{.027} & 0.376\pms{.034} & 12 \\
Gemini 3 Flash  & LLM     & 0.147\pms{.012} & 0.897 & 0.080 & 0.133\pms{.016} & 0.125\pms{.000} & 12 \\
\bottomrule
\end{tabular}
\end{table}

\begin{table}[!ht]
\centering
\caption{Edge type recall by method (aggregated over 3 codebases). $n$ = total ground-truth edges of each type. \textbf{Bold} = best non-Oracle recall per type.}
\label{tab:edge_types}
\begin{tabular}{@{}lcccc@{}}
\toprule
\textbf{Method} & \textsc{imports} & \textsc{calls\_api} & \textsc{data\_flows} & \textsc{reg.\_wires} \\
 & ($n{=}150$) & ($n{=}39$) & ($n{=}16$) & ($n{=}19$) \\
\midrule
Config-Aware    & 0.58 & 0.00 & 0.00 & \textbf{1.00} \\
Random          & 0.55 & 0.00 & 0.00 & 0.00 \\
BFS-Import      & 0.25 & 0.00 & 0.00 & 0.00 \\
\midrule
GPT-5.3-Codex   & \textbf{0.69} & 0.15 & \textbf{0.31} & \textbf{1.00} \\
Claude Sonnet 4.6 & 0.56 & \textbf{0.15} & 0.00 & \textbf{1.00} \\
Gemini 2.5 Flash & 0.47 & \textbf{0.18} & 0.00 & 0.37 \\
Gemini 3.1 Pro  & 0.27 & 0.00 & \textbf{0.50} & 0.00 \\
Gemini 2.5 Pro  & 0.12 & 0.08 & 0.00 & 0.53 \\
Gemini 3 Flash  & 0.11 & 0.00 & 0.00 & 0.05 \\
\bottomrule
\end{tabular}
\end{table}

\begin{figure}[!ht]
    \centering
    \begin{subfigure}[b]{0.48\textwidth}
        \centering
        \includegraphics[width=\textwidth]{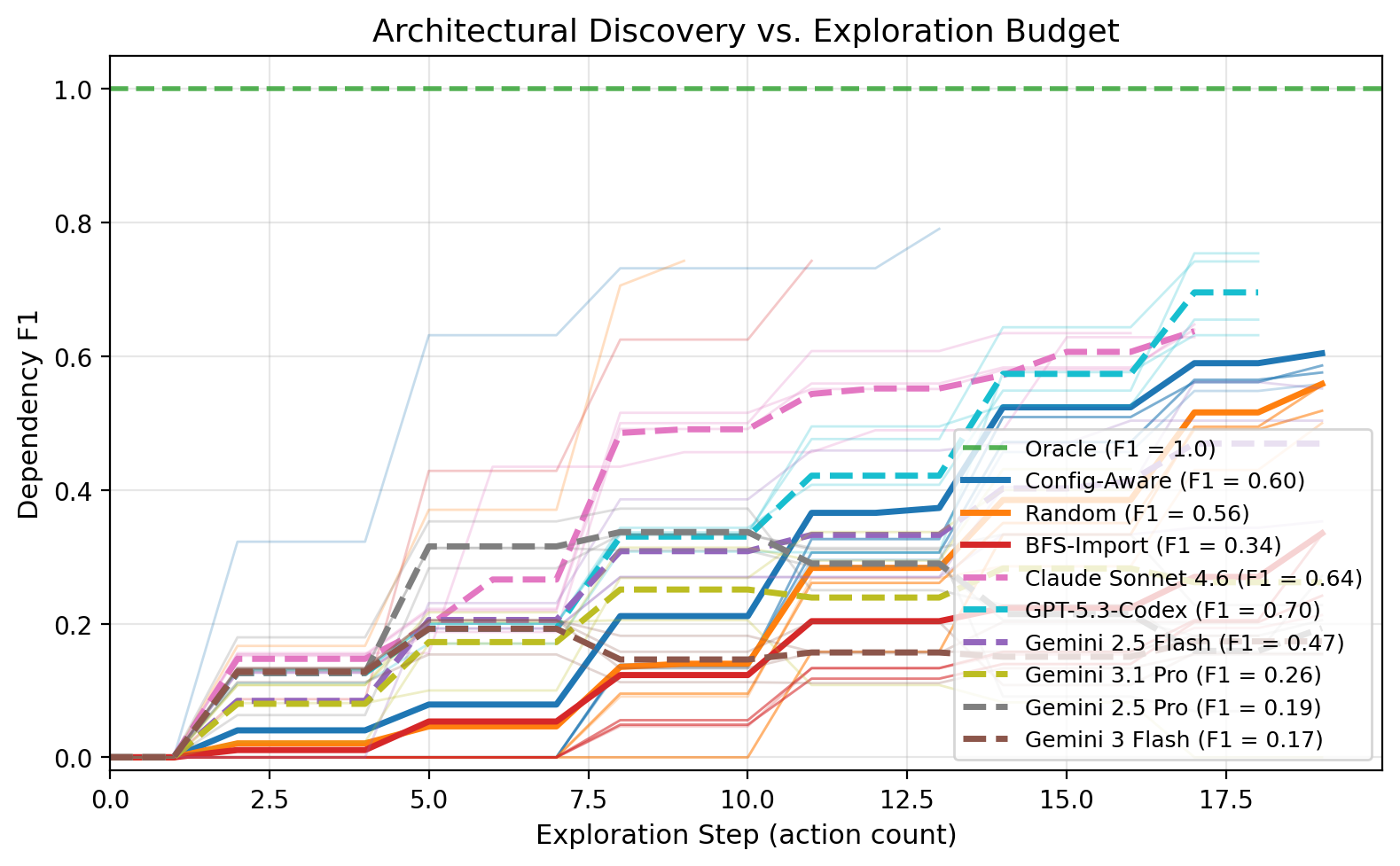}
        \caption{F1 vs.\ exploration steps. Dashed = LLM agents, solid = baselines. GPT-5.3-Codex reaches highest final F1; Claude Sonnet~4.6 leads on efficiency (AUC).}
        \label{fig:f1_steps}
    \end{subfigure}
    \hfill
    \begin{subfigure}[b]{0.48\textwidth}
        \centering
        \includegraphics[width=\textwidth]{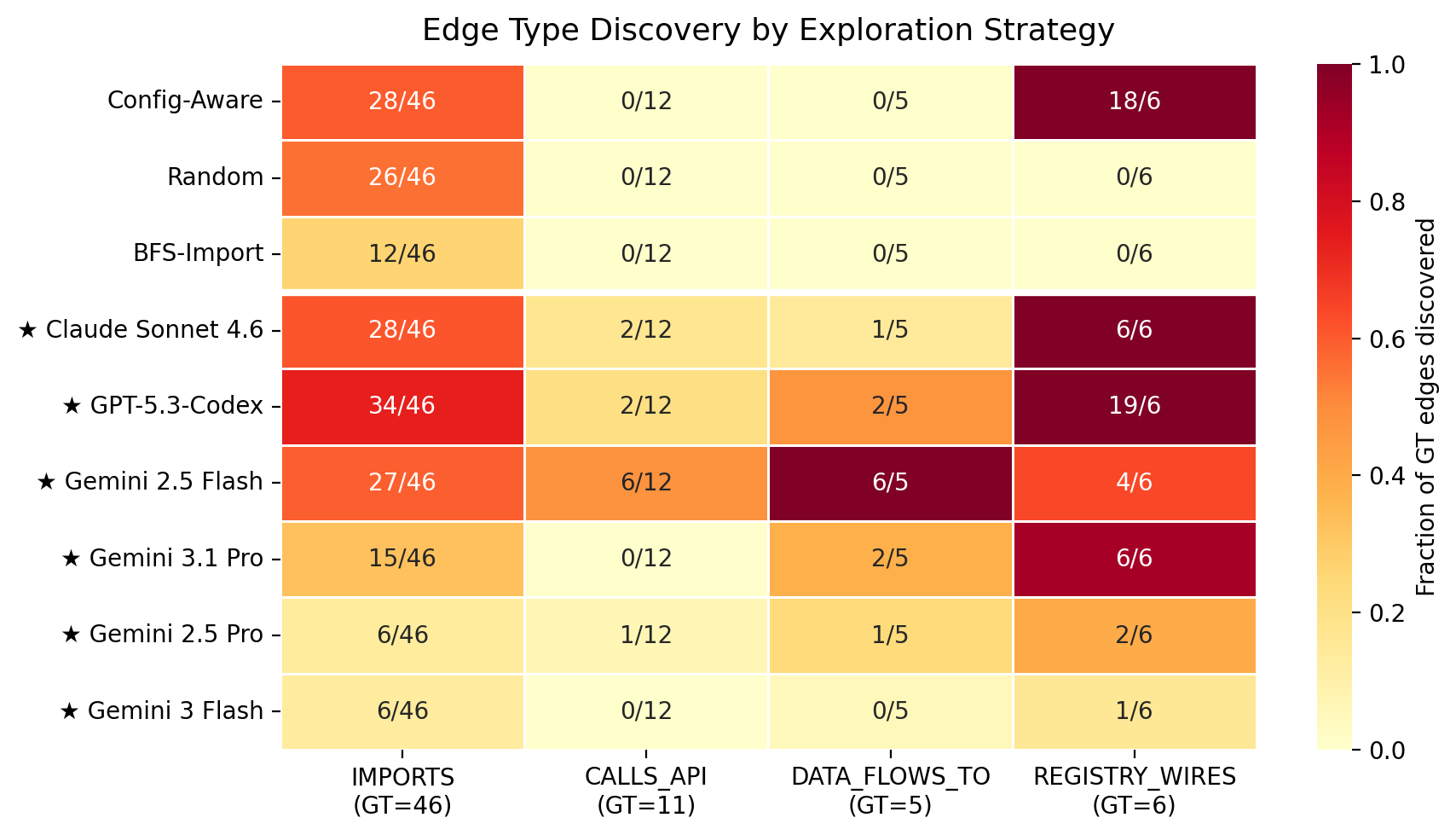}
        \caption{Edge type discovery heatmap. Cells show raw count / ground truth. LLM agents ($\star$) discover all four edge types; baselines find at most two.}
        \label{fig:edge_types}
    \end{subfigure}

    \vspace{0.5em}
    \begin{subfigure}[b]{0.70\textwidth}
        \centering
        \includegraphics[width=\textwidth]{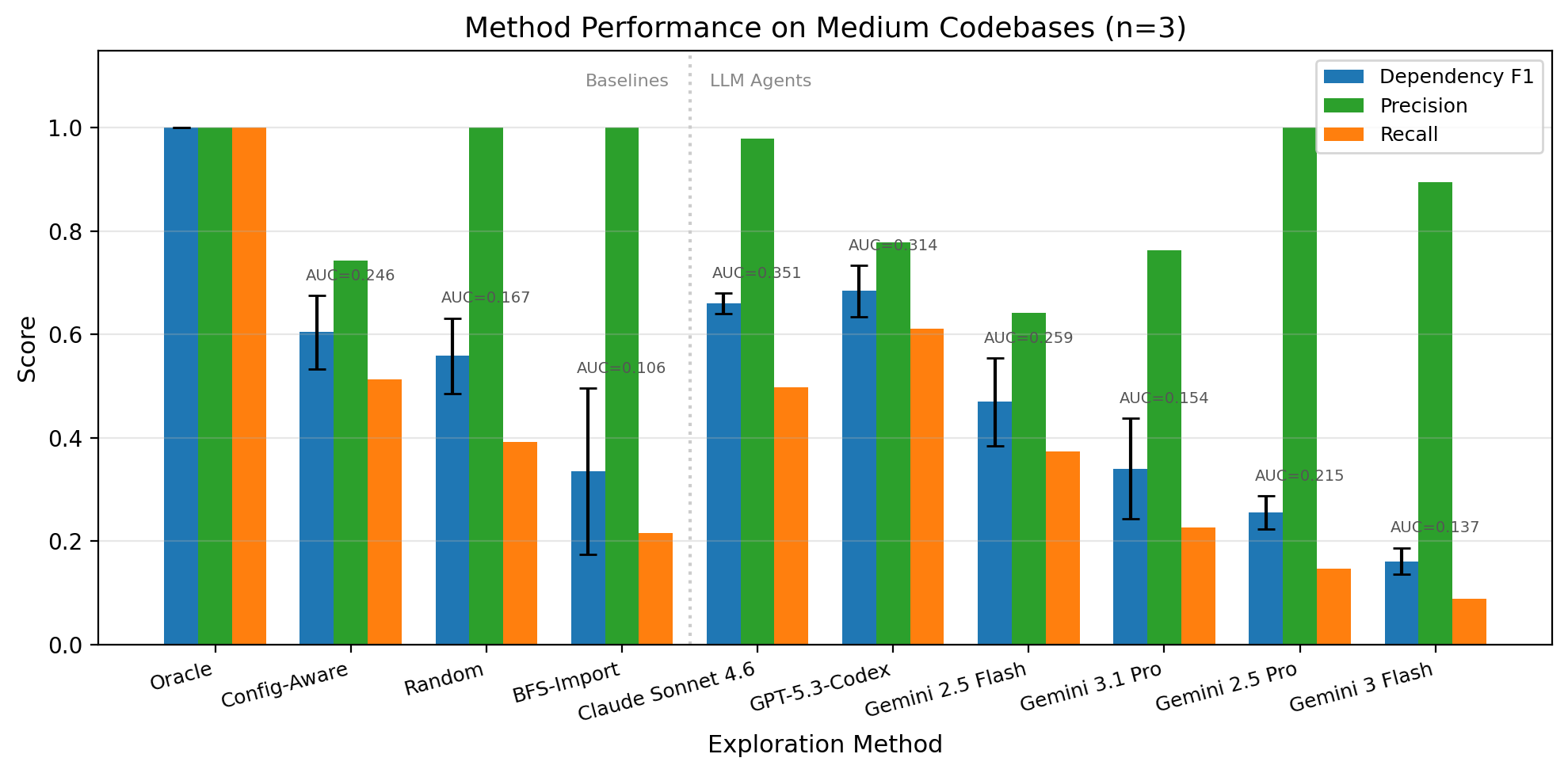}
        \caption{Precision, recall, and F1 per method. GPT-5.3-Codex and Claude Sonnet~4.6 surpass all baselines; weaker LLMs show high precision but low recall.}
        \label{fig:bar_chart}
    \end{subfigure}
    \caption{Exploration analysis across 4 baselines and 6 LLM agents on medium-complexity codebases ($n{=}3$, budget $B{=}20$).}
    \label{fig:main}
\end{figure}

\subsection{Analysis}

\paragraph{Two LLM agents surpass all baselines.}
GPT-5.3-Codex (F1~=~0.676) and Claude Sonnet~4.6 (F1~=~0.664) both exceed Config-Aware (0.577) by 9--10 points. GPT favors broad coverage (highest recall, 0.597); Claude favors targeted exploration (near-perfect precision, 0.983). Both discover all ground-truth REGISTRY\_WIRES edges.

\paragraph{LLM agents discover all four edge types.}
\textbf{LLM agents collectively discover all four edge types} (Table~\ref{tab:edge_types}), while baselines find at most two. The most distinctive capability is DATA\_FLOWS\_TO discovery---requiring multi-hop reasoning through the orchestrator---where Gemini~3.1~Pro achieves 50\% recall and GPT-5.3-Codex 31\%. Claude found zero DATA\_FLOWS\_TO edges despite reading the runner code, illustrating an \emph{exploration-comprehension tradeoff}: understanding the orchestrator is insufficient without also opening the linked stage modules.

\paragraph{Improved prompts unlock invariant discovery.}
With improved probe prompts, LLM agents achieve substantial invariant F1 under relaxed matching: Claude leads (0.778), followed by GPT (0.739) and Gemini~2.5~Flash (0.517). All models scored zero under the earlier prompt version, confirming this was primarily a prompt specification problem (\S\ref{sec:lessons}).

\begin{figure}[!ht]
    \centering
    \includegraphics[width=0.65\textwidth]{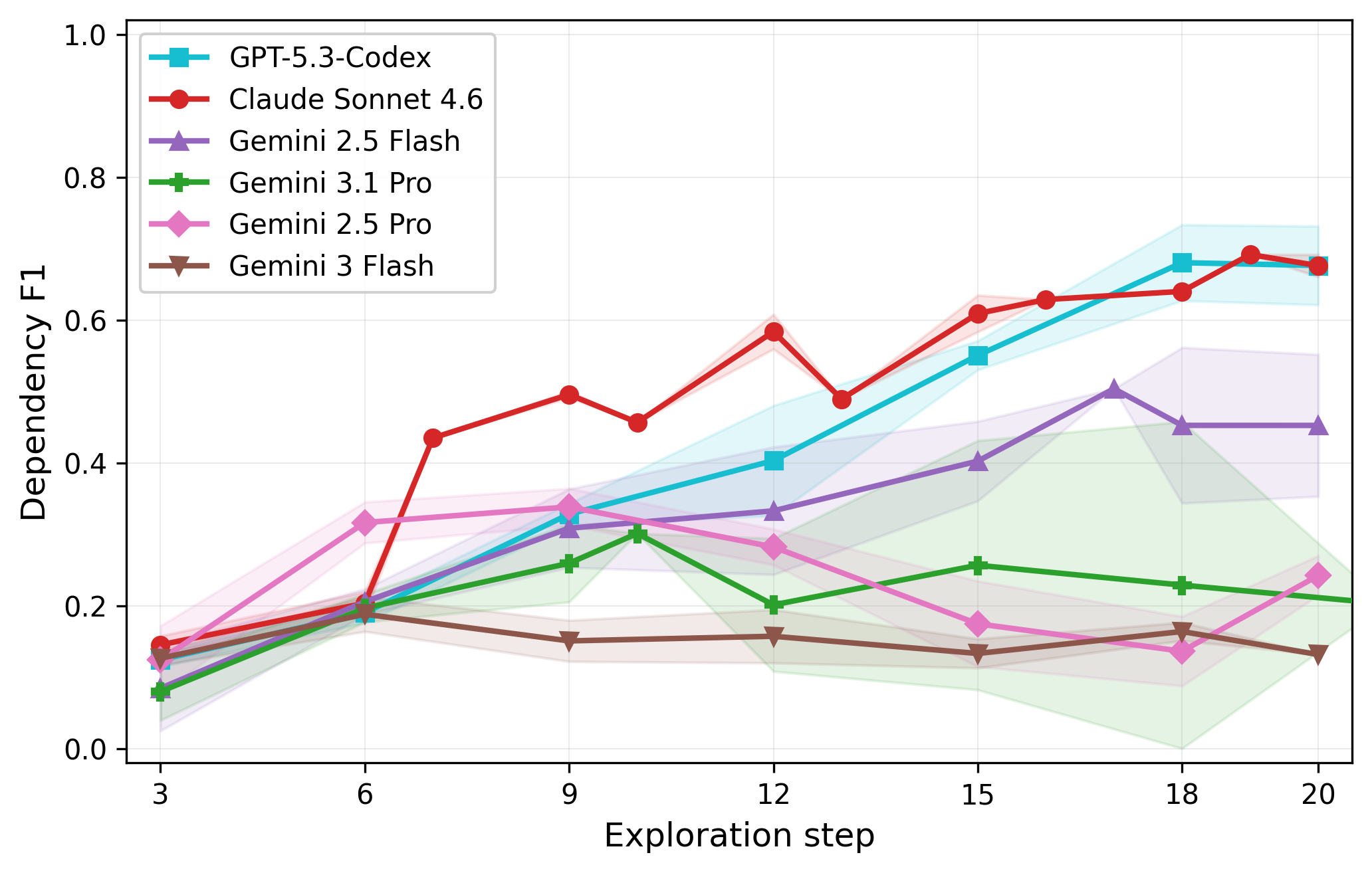}
    \caption{Belief trajectory: Dependency F1 at each probe step (mean over 3 codebases, shaded = std). GPT-5.3-Codex and Claude Sonnet~4.6 accumulate steadily; Gemini~2.5~Flash grows monotonically. Gemini~2.5~Pro peaks at step~6--9 then collapses; Gemini~3.1~Pro shows high variance (stable on one codebase, collapse on another); Gemini~3~Flash remains flat throughout (recency bias).}
    \label{fig:trajectory}
\end{figure}

\paragraph{Belief state instability.}
Gemini~2.5~Pro and 3~Flash reveal catastrophic belief state instability despite opening comparable numbers of files (Figure~\ref{fig:trajectory}). Gemini~3~Flash exhibits recency bias---each probe reports only recently-examined components. Gemini~2.5~Pro builds a reasonable map (peak F1~=~0.33 at step~9), then destroys it in a single probe. In contrast, Gemini~2.5~Flash---the smallest model---loses \textbf{zero} correct edges across all probes. This suggests belief state maintenance is not a function of model scale but may depend on training objectives.

\paragraph{Precision-recall decoupling.}
Claude Sonnet~4.6 and Gemini~2.5~Pro both achieve near-perfect precision ($\geq$0.98), but with dramatically different recall (0.502 vs.\ 0.138)---high precision alone does not indicate strong understanding. GPT trades precision (0.782) for the highest recall (0.597).

\paragraph{Budget effects.}
Random exploration (F1~=~0.538) performs nearly as well as Config-Aware (0.577) at $B{=}20$, but at $B{=}10$ Config-Aware's advantage grows to 3$\times$ (Appendix~\ref{app:budget}). The meaningful discrimination at $B{=}20$ lies in \emph{understanding what you've seen}: LLM agents discover qualitatively different edge types despite comparable file coverage.

\subsection{Active-Passive Gap Decomposition}
\label{sec:apg}

We evaluate the APG decomposition (\S\ref{sec:benchmark}) on two models---GPT-5.3-Codex and Gemini~2.5~Flash---across all three codebases. Table~\ref{tab:apg} reports Dependency F1 under each condition.

\begin{table}[h]
\centering
\caption{Active-Passive Gap decomposition (mean\pms{half-range} over 3 codebases, $B{=}20$). APG$_\text{total}$ = passive-full $-$ active; APG$_\text{selection}$ = oracle $-$ active; APG$_\text{decision}$ = active $-$ replay.}
\label{tab:apg}
\begin{tabular}{@{}llccl@{}}
\toprule
\textbf{Model} & \textbf{Condition} & \textbf{Dep F1} & \textbf{Inv F1} & \textbf{Description} \\
\midrule
\multirow{4}{*}{\rotatebox{90}{\small GPT-5.3}}
& Passive-full    & 0.457\pms{.147} & 0.757\pms{.028} & All files at once \\
& Passive-replay  & 0.665\pms{.034} & 0.752\pms{.036} & Same trace, no decisions \\
& Active          & 0.676\pms{.059} & 0.739\pms{.067} & Agent chooses actions \\
& Passive-oracle  & 0.737\pms{.043} & 0.683\pms{.091} & Oracle-selected files \\
\midrule
\multirow{4}{*}{\rotatebox{90}{\small Gem.~2.5F}}
& Passive-full    & 0.696\pms{.125} & 0.696\pms{.120} & All files at once \\
& Passive-replay  & 0.561\pms{.141} & 0.665\pms{.037} & Same trace, no decisions \\
& Active          & 0.469\pms{.099} & 0.516\pms{.155} & Agent chooses actions \\
& Passive-oracle  & 0.641\pms{.058} & 0.349\pms{.079} & Oracle-selected files \\
\midrule
\multicolumn{5}{@{}l@{}}{\emph{APG decomposition (Dep F1):}} \\
\multicolumn{2}{@{}l}{$\text{APG}_\text{total}$}  & $-$0.219 / $+$0.226 & & GPT: active $>$ full; Gemini: full $>$ active \\
\multicolumn{2}{@{}l}{$\text{APG}_\text{selection}$} & $+$0.060 / $+$0.172 & & Oracle helps both; larger gap for Gemini \\
\multicolumn{2}{@{}l}{$\text{APG}_\text{decision}$}  & $+$0.011 / $-$0.092 & & GPT: $\approx$0; Gemini: passive-replay $>$ active \\
\bottomrule
\end{tabular}
\end{table}

\paragraph{The APG direction is model-dependent.}
In Theory of Space~\citep{zhang2026theoryofspace}, models consistently degrade in active mode ($\text{APG} > 0$). In code, the direction depends on the model. GPT-5.3-Codex exhibits $\text{APG}_\text{total} = -0.219$: active exploration \emph{outperforms} receiving the entire codebase at once, consistent across all three codebases. The likely explanation is \textbf{information overload}: presenting 27--30 files simultaneously overwhelms the model's ability to identify dependency relationships, while sequential exploration allows focused processing. Gemini~2.5~Flash shows the \emph{opposite} pattern ($\text{APG}_\text{total} = +0.226$): passive-full substantially outperforms active (Dep F1 0.696 vs.\ 0.469). This suggests that Gemini benefits from seeing all files at once rather than exploring incrementally---consistent with its strong ``thinking'' reasoning capabilities but weaker exploration strategy.

\paragraph{Selection cost is positive for both models.}
Passive-oracle exceeds Active for both GPT (+0.060) and Gemini (+0.172), indicating suboptimal file selection in both cases. The larger gap for Gemini suggests its exploration strategy is more inefficient, opening less informative files relative to the oracle ordering.

\paragraph{Decision cost reveals a qualitative difference.}
For GPT-5.3-Codex, Active $\approx$ Passive-replay (0.676 vs.\ 0.665): the cognitive overhead of making exploration decisions is negligible. For Gemini~2.5~Flash, the pattern \emph{reverses}: Passive-replay (0.561) \emph{exceeds} Active (0.469) by 9 points ($\text{APG}_\text{decision} = -0.092$). When given GPT's observation trace without having to make decisions, Gemini builds better maps than when exploring on its own. This ``negative decision cost'' suggests that Gemini's exploration decisions actively \emph{hurt} its comprehension---perhaps by directing attention to less informative files or by the cognitive overhead of action selection competing with architectural reasoning.

\paragraph{Invariant discovery patterns.}
For both models, passive-full yields the highest invariant F1 (GPT: 0.757, Gemini: 0.696). Passive-oracle scores lowest for both (GPT: 0.683, Gemini: 0.349). Seeing all files at once provides the global context needed for cross-cutting constraint discovery, while sequential presentation---even with optimal file ordering---limits the model's ability to synthesize invariants spanning multiple modules. Gemini's invariant scores under passive-oracle are notably low (0.349), suggesting it struggles to combine evidence presented incrementally.

\subsection{Probe Condition Ablation}
\label{sec:probe_ablation}

We evaluate the effect of probe conditions on both GPT-5.3-Codex and Gemini~2.5~Flash by comparing three tracking modes: \texttt{probe\_as\_scratchpad} (baseline---JSON map retained in context), \texttt{no\_probe} (final map only, no intermediate probes), and \texttt{probe\_only} (JSON collected but stripped from context after each probe). Table~\ref{tab:probe_ablation} reports results across 3 codebases.

\begin{table}[h]
\centering
\caption{Probe condition ablation (mean\pms{half-range} over 3 codebases, $B{=}20$). Scratchpad effect = scratchpad $-$ no-probe.}
\label{tab:probe_ablation}
\begin{tabular}{@{}llcc@{}}
\toprule
\textbf{Model} & \textbf{Track} & \textbf{Dep F1} & \textbf{Inv F1} \\
\midrule
\multirow{3}{*}{GPT-5.3-Codex}
& Probe-as-scratchpad & 0.676\pms{.059} & 0.739\pms{.067} \\
& No-probe            & 0.538\pms{.127} & 0.570\pms{.191} \\
& Probe-only          & 0.464\pms{.056} & 0.447\pms{.028} \\
\midrule
\multirow{3}{*}{Gemini 2.5 Flash}
& Probe-as-scratchpad & 0.469\pms{.099} & 0.516\pms{.155} \\
& No-probe            & 0.480\pms{.161} & 0.273\pms{.040} \\
& Probe-only          & 0.401\pms{.143} & 0.481\pms{.222} \\
\midrule
\multicolumn{4}{@{}l@{}}{\emph{Scratchpad effect (scratchpad $-$ no-probe):}} \\
\multicolumn{2}{@{}l}{GPT-5.3-Codex} & $+$0.138 & $+$0.169 \\
\multicolumn{2}{@{}l}{Gemini 2.5 Flash} & $-$0.011 & $+$0.243 \\
\bottomrule
\end{tabular}
\end{table}

\paragraph{Scratchpad provides a substantial boost for GPT but not for Gemini.}
For GPT-5.3-Codex, retaining probe JSON in context improves Dep F1 by 13.8 points over the no-probe baseline. The model uses its own prior maps as \textbf{external working memory}, scaffolding subsequent exploration and maintaining belief coherence across probes. This quantitatively confirms the mechanism described in L6 (\S\ref{sec:discussion}): the probe is not a passive measurement but an active intervention that shapes exploration.

For Gemini~2.5~Flash, the scratchpad effect on Dep F1 is negligible ($-$0.011): no-probe (0.480) and scratchpad (0.469) perform comparably. However, scratchpad provides a large boost to \emph{invariant} discovery ($+$0.243), suggesting that Gemini uses the retained map primarily for cross-cutting reasoning rather than dependency tracking. This model-dependent pattern reveals that the scratchpad mechanism works through different channels for different models.

\paragraph{Probe-only is consistently the weakest condition.}
For both models, probe-only yields the lowest Dep F1 (GPT: 0.464, Gemini: 0.401). Producing structured output consumes attention and context budget without the compensating benefit of self-reference. For GPT, probe-only is even below no-probe ($-$0.074), confirming that the overhead of generating structured JSON without retention is actively harmful. For Gemini, the gap is smaller but in the same direction ($-$0.079).

\paragraph{Implications.}
For GPT, the ordering is clear: \textbf{scratchpad $>$ no-probe $>$ probe-only}, with a 14-point F1 gap between scratchpad and no-probe. For Gemini, the ordering is \textbf{no-probe $\approx$ scratchpad $>$ probe-only}, with the primary scratchpad benefit appearing in invariant discovery rather than dependency tracking. This model-dependent pattern complicates benchmark design: the ``best'' probe condition is not universal. We retain probe-as-scratchpad as the default because it enables the richest trajectory analysis and provides the strongest results for the highest-performing model, but report all three conditions for transparency.

% ============================================================================
\section{Discussion}
\label{sec:discussion}

\subsection{Implications for Code Agent Design}

Current code agents (SWE-agent, Aider, Claude Code, Cursor) use tool-augmented retrieval to navigate repositories, but none explicitly construct or maintain a structured architectural belief. Our results show that the strongest LLM agents \emph{can} surpass rule-based strategies when they combine semantic comprehension with reliable belief externalization (GPT-5.3-Codex and Claude Sonnet~4.6 both exceed all baselines), but that weaker models fail at the externalization step despite demonstrating code understanding in their exploration behavior.

This suggests four improvement paths: (1)~\textbf{hybrid approaches} combining AST-level import extraction with LLM semantic analysis, ensuring structural completeness while adding semantic depth; (2)~\textbf{belief externalization training}---explicitly optimizing models to faithfully serialize architectural knowledge into structured formats; (3)~\textbf{exploration strategy optimization}---our APG decomposition (\S\ref{sec:apg}) shows a 6--17-point gap between oracle and agent file selection across models, confirming that \emph{which} files to open is a key bottleneck; and (4)~\textbf{explicit state management}---our probe ablation (\S\ref{sec:probe_ablation}) shows that retaining structured belief maps in context yields a 14-point F1 boost for GPT, empirically validating the value of persistent, accumulative state structures---though the effect is model-dependent, indicating that self-scaffolding is itself a capability that varies across models.

\subsection{The Belief Externalization Problem}

GPT-5.3-Codex recalled 69\% of IMPORTS edges while Gemini~3~Flash and 2.5~Pro recalled only 11--12\% despite opening similar numbers of files. This 6$\times$ gap---for edges the models demonstrably \emph{saw}---shows that belief externalization capability varies dramatically across model families. Both GPT and Claude demonstrate that faithful externalization \emph{is} achievable: the challenge is model-specific, not fundamental.

\paragraph{Belief state instability.}
The Gemini family reveals three distinct instability patterns. Gemini~2.5~Pro shows \emph{catastrophic collapse}: building a map by step~9 (peak F1~=~0.33), then losing 12 correct edges in a single probe. Gemini~3~Flash shows \emph{recency bias}: each probe reports only 3--5 recently-examined components. Gemini~3.1~Pro shows \emph{variable instability}: monotonic accumulation on its best codebase but collapse to zero edges on its worst. Critically, Gemini~2.5~Flash---the smallest model---exhibits \textbf{zero} correct-edge loss across all probes. This suggests belief state maintenance depends on training objectives, not model scale, and may stem from treating each probe as ``summarize from scratch'' rather than ``incrementally update.'' Externalization quality is also likely prompt-dependent; future work should include prompt ablation studies to disentangle model capability from prompt sensitivity.

\paragraph{Error analysis: prompt ambiguity drives false positives.}
Of Gemini~2.5~Flash's 40 false positive edges, \textbf{only 2 (5\%) are true hallucinations}. The dominant failure (45\%) is \emph{edge type confusion}: reporting both IMPORTS and CALLS\_API when code both imports and calls a symbol---arguably reasonable given underspecified definitions. Another 22\% come from treating non-component files as edge endpoints. This reveals that \textbf{prompt specification is a first-order variable}: what looks like an externalization failure may be a legitimate alternative interpretation.

\begin{table}[h]
\centering
\caption{Error analysis: edge type confusion on \texttt{cli.py} edges. Gemini reports both IMPORTS and CALLS\_API for the same target (symbol-qualified), producing false positives; Claude merges into a single file-level edge.}
\label{tab:error_analysis}
\small
\begin{tabular}{@{}lll@{}}
\toprule
\textbf{Edge} & \textbf{Gemini 2.5 Flash} & \textbf{Claude Sonnet 4.6} \\
\midrule
\texttt{cli $\to$ config} & IMPORTS & IMPORTS \\
\texttt{cli $\to$ runner} & IMPORTS & IMPORTS \\
\texttt{cli $\to$ config::PipelineConfig} & CALLS\_API {\color{red}(FP)} & --- \\
\texttt{cli $\to$ runner::run\_pipeline} & CALLS\_API {\color{red}(FP)} & --- \\
\texttt{cli $\to$ runner} & --- & CALLS\_API {\color{green!50!black}(TP)} \\
\bottomrule
\end{tabular}
\end{table}

\subsection{Benchmark Design Lessons}
\label{sec:lessons}

Building \tocs{} v0.1 exposed several pitfalls in belief-probing benchmark design:

\textbf{L1: Edge type decision rules.} When code both imports and calls a symbol, the prompt must specify which edge type to report. Ambiguity here caused 45\% of Gemini~2.5~Flash's false positives. \emph{Fix:} explicit decision trees.

\textbf{L2: Component boundaries.} Without defining whether test files or configs count as ``components,'' 22\% of FPs came from including non-component endpoints. \emph{Fix:} explicit inclusion rules in the probe.

\textbf{L3: Invariant field semantics.} Generic fields without per-type definitions yielded Inv F1~=~0.0 for all models. Adding worked examples jumped scores to 0.52--0.78 under relaxed matching.

\textbf{L4: Alternative correct answers.} Strict exact-match scoring penalizes semantically correct but differently-formatted answers. \emph{Fix:} tiered or normalized scoring.

\textbf{L5: Prompt specification as experimental variable.} What looks like a model capability gap may be a prompt specification gap. A single prompt change moved Gemini~2.5~Flash's F1 by $+$0.141 while barely affecting Claude ($+$0.012; Table~\ref{tab:prompt_ablation} in Appendix).

\textbf{L6: Probes as active interventions.} In scratchpad mode, improved probe prompts change not just serialization quality but \emph{exploration behavior}---on one codebase, GPT-5.3-Codex opened 4 additional stage modules after a more structured probe revealed gaps in its own map (F1: 0.392~$\to$~0.636). The probe condition ablation (\S\ref{sec:probe_ablation}) quantifies this: scratchpad boosts GPT by 14 F1 points but has no dependency-level effect on Gemini---suggesting different models leverage self-scaffolding through different channels.

\subsection{Limitations}
\label{sec:limitations}

\tocs{} v0.1 evaluates a single architectural pattern (Pipeline) in a single language (Python) with neutral filenames (\texttt{mod\_a.py}), no standalone documentation, and no organic complexity. Six models from three providers are each evaluated on 3 codebases with a single run per model-codebase pair. All results use a single prompt design; different formulations could yield substantially different rankings. The cognitive map externalization is a proxy for internal representation---a model may ``know'' more than it serializes.

\subsection{Roadmap}
\label{sec:roadmap}

Phase~1 (current): benchmark construction---stress-test the framework with diverse models. Phase~2: community-contributed codebase patterns, additional languages, documentation as a discovery channel, and REVISE mode evaluation. Phase~3: large-scale controlled experiments with repeated runs, open-weight models, and prompt ablation studies. We release the complete toolkit as open-source and invite contributions.

% ============================================================================
\section{Conclusion}
\label{sec:conclusion}

We introduced \tocs{}, the first benchmark for active architectural belief construction in code. Three findings stand out. First, the Active-Passive Gap is model-dependent: GPT-5.3-Codex builds better maps through active exploration ($\text{APG} = -0.22$) while Gemini~2.5~Flash does better receiving all files at once ($\text{APG} = +0.23$)---showing that active exploration is itself a non-trivial capability. Second, self-scaffolding through belief externalization is model-dependent: scratchpad mode boosts GPT by 14 F1 points but not Gemini. Third, belief state maintenance varies dramatically: the smallest Gemini model maintains perfectly stable beliefs while its larger sibling suffers catastrophic collapse. These findings validate the benchmark's design---periodic probing reveals dynamics invisible to final-snapshot evaluation---while the design lessons learned from building it (\S\ref{sec:lessons}) are themselves a contribution. We release \tocs{} as an open benchmark and invite community contributions.

% ============================================================================
\bibliographystyle{plainnat}
\bibliography{references}

% ============================================================================
\appendix

\section{Evaluation Prompts}
\label{app:prompts}

\paragraph{System prompt (excerpt).} The agent is told: ``You are an expert software engineer exploring a codebase you have never seen before. You are operating under PARTIAL OBSERVABILITY [...]  Available actions: LIST, OPEN, SEARCH, INSPECT, DONE. [...] Your goal is to build a complete and accurate ARCHITECTURAL UNDERSTANDING.'' The full prompt describes action semantics and exploration strategy tips.

\paragraph{Cognitive map probe.} After every $K$ actions, the agent receives: ``Externalize your current architectural belief as JSON matching the following schema [...]'' The schema specifies components with status/purpose/exports/edges/confidence, invariants with structured canonical form, and an uncertainty summary. A worked example demonstrates the expected format (see supplementary materials).

\section{Generated Codebase Example}
\label{app:example}

A medium-complexity codebase (seed 42, text processing domain) generates the following structure:
\begin{verbatim}
text_processor/
+-- __init__.py          +-- middleware/
+-- models.py            |   +-- mod_i.py (logging)
+-- base.py              |   +-- mod_j.py (retry)
+-- config.py            +-- utils/
+-- exceptions.py        |   +-- helpers.py
+-- registry.py          |   +-- formatters.py
+-- stages/              +-- legacy/
|   +-- mod_a.py         |   +-- old_pipeline.py
|   +-- mod_b.py         |   +-- compat.py
|   +-- mod_c.py         +-- runner.py
|   +-- mod_d.py         +-- cli.py
|   +-- mod_e.py         +-- pipeline_config.json
|   +-- mod_f.py         +-- test_smoke.py
+-- adapters/
|   +-- mod_g.py
|   +-- mod_h.py
\end{verbatim}

Stages implement a \texttt{StageBase} ABC. The registry loads stage ordering from \texttt{pipeline\_config.json}. Stages cannot import each other directly (enforced by test). Legacy modules are distractors not in the pipeline flow.

\section{Budget Sweep}
\label{app:budget}

To characterize how exploration budget affects discrimination between strategies, we ran all four baselines at $B \in \{10, 15, 20, 25\}$ across 3 codebases. Table~\ref{tab:budget} reports mean Dependency F1.

\begin{table}[h]
\centering
\caption{Dependency F1 vs.\ exploration budget (mean over 3 codebases).}
\label{tab:budget}
\begin{tabular}{@{}lcccc@{}}
\toprule
\textbf{Method} & $B{=}10$ & $B{=}15$ & $B{=}20$ & $B{=}25$ \\
\midrule
Oracle       & 1.000 & 1.000 & 1.000 & 1.000 \\
Config-Aware & 0.175 & 0.492 & 0.577 & 0.626 \\
Random       & 0.056 & 0.317 & 0.538 & 0.632 \\
BFS-Import   & 0.078 & 0.157 & 0.293 & 0.603 \\
\bottomrule
\end{tabular}
\end{table}

At tight budgets ($B{=}10$), Config-Aware's strategy of prioritizing registry files yields a 3$\times$ advantage over Random (0.175 vs.\ 0.056). As budget increases, raw coverage dominates: at $B{=}25$, Random matches Config-Aware. The meaningful discrimination at $B{=}20$ lies in semantic edge discovery by LLM agents.

\section{Prompt Sensitivity}
\label{app:prompt_sensitivity}

\begin{table}[h]
\centering
\caption{Effect of improved probe prompt on Dependency F1 (same 3 codebases, same models, only probe prompt changed). $\Delta$ = new $-$ old.}
\label{tab:prompt_ablation}
\begin{tabular}{@{}lccc@{}}
\toprule
\textbf{Model} & \textbf{Old Prompt} & \textbf{New Prompt} & $\boldsymbol{\Delta}$\textbf{F1} \\
\midrule
Gemini 2.5 Flash & 0.328 & 0.469 & +0.141 \\
GPT-5.3-Codex & 0.564 & 0.676 & +0.113 \\
Claude Sonnet 4.6 & 0.639 & 0.664 & +0.025 \\
\bottomrule
\end{tabular}
\end{table}

Models with weaker baseline externalization benefit disproportionately from prompt improvements. Claude Sonnet~4.6, which already externalized comprehensively under the old prompt, showed minimal change ($+$0.025). Main-text results use the improved prompt throughout.

\end{document}